\documentclass[longauth]{aa}  

\usepackage{graphicx}
\usepackage{txfonts}
\usepackage{calrsfs}

\def\beq{\begin{equation}}
\def\eeq{\end{equation}}
\usepackage{calrsfs}
\DeclareMathAlphabet{\pazocal}{OMS}{zplm}{m}{n}

\usepackage{xcolor}

\begin{document}

   \title{Exploring the presence of a fifth force \\ at the Galactic Center}

\author{
    The GRAVITY Collaboration\fnmsep\thanks{
    GRAVITY is developed in collaboration by MPE, LESIA of Paris Observatory / CNRS / Sorbonne Université / Univ. Paris Diderot, and IPAG of Université Grenoble Alpes / CNRS, MPIA, Univ. of Cologne, CENTRA - Centro de Astrofisica e Gravitação, and ESO. } 
    : 
    K.~Abd El Dayem             \inst{1}            \and
    R.~Abuter                   \inst{4}            \and
    N.~Aimar                    \inst{10, 7}            \and
    P.~Amaro Seoane             \inst{14,2,19}   \and
    A.~Amorim                   \inst{8,7}          \and
    J.P.~Berger                 \inst{3,4}          \and 
    H.~Bonnet                   \inst{4}            \and
    G.~Bourdarot                \inst{2}            \and
    W.~Brandner                 \inst{5}            \and
    V.~Cardoso                  \inst{7,16}         \and
    Y.~Clénet                   \inst{1}            \and
    R.~Davies                   \inst{2}            \and
    P.T.~de~Zeeuw               \inst{20}           \and
    A.~Drescher                 \inst{2}            \and
    A.~Eckart                   \inst{6,13}         \and
    F.~Eisenhauer               \inst{2,18}         \and
    H.~Feuchtgruber             \inst{2}            \and
    G.~Finger                   \inst{2}            \and
    N.M.~Förster~Schreiber      \inst{2}            \and
    A.~Foschi                   \inst{1, 2}\thanks{Corresponding author: A.~Foschi (arianna.foschi{@}obspm.fr).}          \and
    P.~Garcia                   \inst{10,7}         \and
    E.~Gendron                  \inst{1}            \and
    R.~Genzel                   \inst{2,11}         \and
    S.~Gillessen                \inst{2}            \and
    M.~Hartl                    \inst{2}            \and
    X.~Haubois                  \inst{9}            \and
    F.~Haussmann                \inst{2}            \and
    T.~Henning                  \inst{5}            \and
    S.~Hippler                  \inst{5}            \and
    M.~Horrobin                 \inst{6}            \and
    L.~Jochum                   \inst{9}            \and
    L.~Jocou                    \inst{3}            \and
    A.~Kaufer                   \inst{9}            \and
    P.~Kervella                 \inst{1}            \and
    S.~Lacour                   \inst{1,4}            \and
    V.~Lapeyrère                \inst{1}            \and
    J.-B.~Le~Bouquin            \inst{3}            \and
    P.~Léna                     \inst{1}            \and
    D.~Lutz                     \inst{2}            \and
    F.~Mang                     \inst{2}            \and
    N.~More                     \inst{2}            \and
    J.~Osorno                   \inst{1}            \and
    T.~Ott                      \inst{2}            \and
    T.~Paumard                  \inst{1}            \and
    K.~Perraut                  \inst{3}            \and
    G.~Perrin                   \inst{1}            \and
    S.~Rabien                   \inst{2}            \and
    D.~C.~Ribeiro               \inst{2}            \and
    M.~Sadun Bordoni            \inst{2}            \and
    S.~Scheithauer              \inst{5}            \and
    J.~Shangguan                \inst{21}            \and
    T.~Shimizu                  \inst{2}            \and
    J.~Stadler                  \inst{12,2}         \and
    O.~Straub                   \inst{2,17}         \and
    C.~Straubmeier              \inst{6}            \and
    E.~Sturm                    \inst{2}            \and
    L.J.~Tacconi                \inst{2}            \and
    I.~Urso                     \inst{1}            \and
    F.~Vincent                  \inst{1}            \and
    S.D.~von~Fellenberg           \inst{13,2}         \and
    E.~Wieprecht                \inst{2}            \and
    J.~Woillez                  \inst{4}            
}

    \institute{
            LIRA, Observatoire de Paris, Université PSL, CNRS, Sorbonne Université, Université de Paris, 5 place Jules Janssen, 92195 Meudon, France \and 
            Max Planck Institute for Extraterrestrial Physics, Giessenbachstraße 1, 85748 Garching, Germany \and 
            Univ. Grenoble Alpes, CNRS, IPAG, 38000 Grenoble, France \and
            European Southern Observatory, Karl-Schwarzschild-Straße 2, 85748 Garching, Germany \and
            Max Planck Institute for Astronomy, Königstuhl 17, 69117 Heidelberg, Germany \and
            1st Institute of Physics, University of Cologne, Zülpicher Straße 77, 50937 Cologne, Germany \and
            CENTRA - Centro de Astrofísica e Gravitação, IST, Universidade de Lisboa, 1049-001 Lisboa, Portugal \and
            Universidade de Lisboa - Faculdade de Ciências, Campo Grande, 1749-016 Lisboa, Portugal \and
            European Southern Observatory, Casilla 19001, Santiago 19, Chile \and
            Faculdade de Engenharia, Universidade do Porto, rua Dr. Roberto Frias, 4200-465 Porto, Portugal \and
            Departments of Physics \& Astronomy, Le Conte Hall, University of California, Berkeley, CA 94720, USA \and
            Max Planck Institute for Astrophysics, Karl-Schwarzschild-Straße 1, 85748 Garching, Germany \and
            Max Planck Institute for Radio Astronomy, auf dem Hügel 69, 53121 Bonn, Germany \and
            Institute of Multidisciplinary Mathematics, Universitat Politècnica de València, València, Spain \and
            Advanced Concepts Team, ESA, TEC-SF, ESTEC, Keplerlaan 1, 2201 AZ Noordwijk, The Netherlands \and
            Center of Gravity, Niels Bohr Institute, Blegdamsvej 17, 2100 Copenhagen, Denmark \and
            ORIGINS Excellence Cluster, Boltzmannstraße 2, 85748 Garching, Germany \and
            Department of Physics, Technical University of Munich, 85748 Garching, Germany \and
            Higgs Centre for Theoretical Physics, Edinburgh, UK \and
            Leiden University, 2311 EZ Leiden, The Netherlands \and
            The Kavli Institute for Astronomy and Astrophysics, Peking University, Beijing 100871, China
    }  

   \date{
   }

 
  \abstract
   {}
   {We investigate the presence of a Yukawa-like correction to Newtonian gravity  at the Galactic Center, leading to a new upper limit on the intensity of such a correction.}
   {We performed a Markov chain Monte Carlo (MCMC) analysis using the astrometric and spectroscopic data of star S$2$ collected at the Very Large Telescope by GRAVITY, NACO, and SINFONI instruments, covering the period from $1992$ to $2022$.}
   {The precision of the GRAVITY instrument allows us to derive the most stringent upper limit at the Galactic Center for the intensity of the Yukawa contribution ($\propto \, \alpha e^{- \lambda r}$) of  $|\alpha| < 0.003$ for a scale length of $\lambda = 3 \cdot 10^{13}\, \rm m\,  (\sim 200 \, \rm  AU)$. This is an improvement on all estimates obtained in previous works  by roughly one order of magnitude. }
   {}


   \maketitle
%

\section{Introduction}
General relativity (GR) is the most widely recognized theory of gravity today. Its predictions have been extensively tested on Solar System scales and using gravitational waves emission by black holes (BHs) and binary pulsars \citep{Will:2014kxa, Will:2018bme, Nitz:2021uxj}. Until now, no significant deviation from GR has been detected in any of these observations.  However, it is also known that beyond the regime we are  currently able to test  experimentally, GR is generally considered insufficient. 

Cosmological observations indicate an expanding Universe whose acceleration can only be explained by introducing a cosmological constant ad hoc, which  lacks a solid theoretical explanation and raises several issues~\citep{Weinberg:1988cp, Peebles:2002gy}. Other observational evidence, such as the rotational curve of galaxies~\citep{1985ApJ...295..305V, Salucci:2018hqu} or gravitational lensing effects \citep{Massey:2010hh} indicate the presence of a dark massive component of the Universe, whose nature is still unknown. Furthermore, it is well known that GR lacks a quantum description at high-energy scales and several attempts have been made to create a theory valid at all scales (for a review on the state-of-the-art approaches, see, e.g., ~\cite{2011arXiv1108.3269E, 2023arXiv230213047K}).

One way to address these inconsistencies between theory and experiments is to directly modify GR, giving rise to a plethora of possible extended theories of gravity (ETG).
In particular, a Yukawa-like interaction emerges quite naturally in the weak field limit of several ETGs; for instance, scalar-tensor-vector theories \citep{Moffat:2005si}, massive gravity theories \citep{Visser:1997hd, Hinterbichler:2011tt}, theories in higher dimensions with Kaluza-Klein compactification \citep{PhysRevLett.57.25, Hoyle:2000cv}, massive Brans-Dicke theories \citep{Perivolaropoulos:2009ak, Alsing:2011er}, or $f(R)$ theories \citep{Capozziello:2015lza}. However, the so-called "fifth-force" scenario has also made an appearance  in a number of specific dark matter models   \citep{PhysRevLett.67.2926, Gradwohl:1992ue, Carroll:2008ub}.

Due to the importance that a modification of Newtonian gravity would have on our understanding of the Universe, the presence of a Yukawa-like contribution has been repeatedly investigated in the past. The fifth-force intensity is well constrained at Solar System scales via the motion of planets \citep{Konopliv:2011, Hees:2014kta, Berge:2017xws, Will:2018gku, 2022GReGr..54...44S}, from the Lunar Laser Ranging experiment \citep{Hofmann:2018} and also by making use of the planetary ephemerides \citep{Mariani:2023ubf, Fienga:2023ocw}. Recent constraints have been obtained from asteroids tracking \citep{Tsai:2021irw, Tsai:2023zza} and test of the weak equivalence principle \citep{MICROSCOPE:2022doy}.

The discovery of orbiting stars around the Galactic Center (GC) \protect\citep{Eckart:1996zz, 2002Natur.419..694S, Ghez_2003, 2009ApJ...692.1075G, 2009ApJ...707L.114G, Sabha:2012vc}, all located within one arcsecond distance from the supermassive black hole (SMBH) Sagittarius A$^*$ (Sgr A$^*$), allows us to test GR in a completely different environment from the Solar System.

The importance of looking for a fifth force in the GC lies in the fact that many ETGs that predict a Yukawa-like term also display a screening mechanism that suppresses the fifth force contribution at Solar System scales and prevents its detection. This would explain why it would be yet unobserved, while its effect may be different around SMBHs. 

The current constraints on the intensity of a fifth force in the GC come from the analysis of SgrA$^*$'s shadow by the Event Horizon Telescope \citep{Vagnozzi:2022moj}, from the measurement of the Schwarzschild precession in S$2$ motion \citep{GRAVITY:2020gka, Jovanovic:2022twh, Jovanovic:2023tcc, Jovanovic:2024kwh} and from the analysis of S-stars publicly available (or mock) data \citep{Borka:2013dba, Capozziello:2014rva,  Borka:2021omc, Zakharov:2016lzv, Zakharov:2018cbj, deMartino:2021daj, DellaMonica:2021xcf}, while also including the presence of a (expected) bulk mass distribution around Sgr A$^*$ \citep{Jovanovic:2021hrz}. For a complete and comprehensive review on astrophysical and theoretical constraints at the GC, we refer to \cite{deLaurentis:2022oqa}. 

In the context of S-stars, a previous work by \cite{Hees:2017aal}, with a full analysis of the S$2$ data, showed that the intensity of such a contribution cannot exceed $\alpha \sim  0.01$ at scales comparable to the S$2$-SgrA$^*$ distance. 

In this Letter, we use the astrometric and spectroscopic measurements of the star S$2$ collected at the Very Large Telescope (VLT) by GRAVITY, NACO, and SINFONI to constrain the intensity of a possible Yukawa correction at the GC. 
Although we do not expect our results to consistently deviate from the estimates obtained in the aforementioned literature, a complete analysis of S$2$ including GRAVITY data, which dominate the $\chi^2$ due to their very small uncertainties, is still lacking. As  we show here, the precision of the GRAVITY instrument allows us to place a significantly stronger constraint than the previous estimates.

\section{Observations}
The set of available data, $D,$ can be organized according to the following criteria:\ 
\begin{itemize}
\item[a)] Astrometric data $\rm DEC$, $\rm R.A.$ 
    \begin{itemize}
        \item 128 data points collected using both the SHARP camera at the New Technology Telescope between 1992 and 2002 ($\sim$ 10 data points, accuracy of $\approx 4 \, \rm mas$) and the NACO imager at the VLT between 2002 and 2019 (118 data points, accuracy of $\approx 0.5 \, \rm mas$);
    
        \item 76 data points collected by GRAVITY at the VLT between 2016 and April 2022 (accuracy of $\approx 50 \, \rm \mu as$).
    \end{itemize}
    
\item[b)] Spectroscopic data $V_R$
    \begin{itemize}
        \item 102 data points collected by SINFONI at the VLT (100 points) and NIRC2 at Keck (2 points) collected between 2000 and March 2022 (accuracy in good conditions of $\approx 10-15 \, \rm km/s $).
    \end{itemize}
\end{itemize}

\section{Yukawa correction to Newtonian force}
\subsection{Model}

The potential we aim to test takes the following form:
\beq
U = -\frac{G M}{r} \left(1 + |\alpha| e^{-r/\lambda}\right)\, , 
\label{yukawa}
\eeq
where $\alpha$ represents the strength of interaction and $\lambda$ is a scale parameter, which depends on the specific theory considered. For example, when new massive fields are included in the theory, $\lambda$ represents the Compton wavelength of the field, which is related to the mass by $m_{\varphi} = h/c \lambda$, where $h$ is the Planck constant. 

In comparison to previous works, here the Schwarzschild precession in the S$2$ orbit is also included, since it has been formally detected at a $10\sigma$ confidence level by the GRAVITY Collaboration \citep{GRAVITY:2020gka, GRAVITY:2024tth}. 

Although a formal parametrized post-Newtonian (PN) treatment is not possible when a massive field is included in the action \citep{Alsing:2011er, PoissonWill2012}, we can still derive the equations of motion of a test particle, assuming the parametrized PN parameters to be $\gamma = \beta = 1$. This latter assumption is valid for ETGs that are indistinguishable from GR at $1$PN order and it is supported by different experimental observations, including those carried out for the GC (see, e.g., \cite{Will:2018bme, Hofmann:2018, GRAVITY:2020gka, Iorio_2024}). On the other hand, non-metric theories of gravity or specific subclasses of f(R) and scalar-tensor theories, with parametrized PN parameters that significantly stray from unity, are excluded from this analysis \citep{Will:2014kxa}.

The total acceleration experienced by the star is
\beq
\mathbf{a}_{\rm TOT} = \mathbf{a}_{\rm New} + \mathbf{a}_{\rm Yuk} + \mathbf{a}_{\rm 1PN}\, ,
\eeq
where $\mathbf{a}_{\rm New} + \mathbf{a}_{\rm Yuk}$ are derived from the potential in Eq.~\eqref{yukawa} and 
\beq
\boldsymbol{a}_{1 \rm PN} = \frac{ G M}{c^2 r^2} \left[\left(\frac{4 G M}{r} - v^2\right) \frac{\boldsymbol{r}}{r} + 4 \dot{r}\boldsymbol{v}  \right] \, ,
\label{1pn}
\eeq 
with $\boldsymbol{r} = r \hat{r}$, $
\boldsymbol{v} = \left(\dot{r} \hat{r}, r \dot{\theta}\hat{\theta}, r \dot{\phi} \sin \theta \hat{\phi} \right)$ and $ v = |\boldsymbol{v}|$.

The above expression coincides with the $1$PN acceleration derived in \cite{Alves:2023cuo} for the two-body problem in massive Brans-Dicke theory.
Section \ref{subsec:comaprison_literature} will be devoted to a comparison of Eq.~\eqref{1pn} with results developed in the literature when extra massive degrees of freedom are included in the theory.  

\subsection{Method}
The numerical integration of the equations of motion was performed using a Runge-Kutta $4(5)$ method, with further details reported in Appendix \ref{app:num_integration}. During the fit of the S$2$ data, the Yukawa length scale, $\lambda$, was kept fixed, with values set between $10^{12} \leq \lambda \leq 10^{15} \, \rm m$, while the intensity, $\alpha$, was allowed to vary together with other parameters describing the system. 

Specifically, the set of parameters is given by:
\begin{equation}
    \Theta_i = \{e, a_{\rm sma}, \Omega_{\rm orb}, i_{\rm orb}, \omega_{\rm orb}, t_p, R_0, M, x_0, y_0, v_{x_0}, v_{y_0}, v_{z_0}, \alpha \},
    \label{emcee_parameters_yukawa}
\end{equation}
where $e$ is the eccentricity and $a_{\rm sma}$ the semi major axis of the star S$2$, while $\Omega_{\rm orb}$, $i_{\rm orb}$, and $\omega_{\rm orb}$ are the three angles used to project the star's orbital frame into the observer reference frame using the procedure reported in Appendix \ref{app:coord_transf}. Then, $t_p$ is the time of pericenter passage, while $M$ and $R_0$ are the SMBH mass and the GC distance, respectively. The additional parameters $\{x_0, y_0,v_{x_0}, v_{y_0}, v_{z_0} \}$ characterize the NACO/SINFONI data reference frame with respect to Sgr~A$^*$ \citep{2015MNRAS.453.3234P}.

To fit the S$2$ data, we perform a Markov chain Monte Carlo (MCMC) analysis using the Python package \textsc{emcee} \citep{2013PASP..125..306F}. The log-likelihood is given by
\begin{equation}
    \ln \mathcal{L} = \ln \mathcal{L}_{\rm pos} + \ln \mathcal{L}_{\rm vel}\,,
\end{equation}
where 
\begin{equation}
    \ln \mathcal{L}_{\rm pos} = - \sum_{i=1}^{N} \left[ \frac{ (\rm DEC_{i} - \rm DEC_{\rm model, i})^2}{\sigma_{\rm DEC_{i}}^2}  +  \frac{ (\rm R.A._{i} - \rm R.A._{\rm model, i})^2}{\sigma_{\rm R.A._{i}}^2} \right]\,,
\end{equation}
and 
\begin{equation}
    \ln \mathcal{L}_{\rm vel} = - \sum_{i=1}^{N} \frac{ (V_{R, i} - V_{\rm model, i})^2}{\sigma_{V_{R, i}}^2} \, .
\end{equation}
The priors are listed in Tables~\ref{table:priors} and \ref{priors_offset}. Uniform priors were used for the physical parameters, that is, we only imposed physically motivated bounds, while Gaussian priors were implemented for the offset parameters, since the latter have been well constrained by an independent previous work and were not expected to change~\citep{2015MNRAS.453.3234P}. The initial points $\Theta_i^0$ in the MCMC are chosen to be the most recent best fit parameters of S$2$ orbit reported in the literature \citep{GRAVITY:2024tth}.

\begin{table}
\caption{Uniform priors used in the MCMC analysis. } 
\label{table:priors}
\begin{tabular}{cccc}
    \hline
    Parameter & $\Theta_i^0$ & Lower bound & Upper bound \\
    \hline
    $e$ & 0.88441 & 0.83 & 0.93 \\[2pt] 
    $a_{\rm sma}$ [''] & 0.12497 & 0.119 & 0.132 \\[2pt]  
    $i_{\rm orb} \, [^\circ] $  & $134.69241$ & 100 & $150$ \\ [2pt] 
    $\omega_{\rm orb} \, [^\circ]$ & $66.28411$ & 40 & $90$ \\ [2pt] 
    $\Omega_{\rm orb} \, [^\circ]$ & $228.19245$ & $200$ & $250$ \\ [2pt] 
    $t_p $ [yrs] & 2018.37902 & 2018 & 2019 \\ [2pt] 
    $M \, [10^6 \, M_{\odot}] $ & $4.29950$ & 4.1 & 4.8\\[2pt] 
    $ R_0 \, \rm [10^3 \, pc]$ & 8.27795 & 8.1 & 8.9\\ [2pt] 
   \hline
\end{tabular}
\end{table}
\begin{table}
\caption{Gaussian priors used in the MCMC analysis.} 
\begin{tabular}{cccc}
    \hline
    Parameter & $\Theta_i^0$ & $\xi$ & $\sigma$ \\
    \hline
    $ x_0 \, \rm [mas]$ & -0.244 & -0.055 & 0.25 \\ [2pt] 
    $ y_0 \, \rm [mas]$ & -0.618 & -0.570 & 0.15 \\ [2pt] 
    $ v_{x_0} \, \rm [mas/yr]$ & 0.059 & 0.063 & 0.0066 \\ [2pt] 
    $ v_{y_0} \, \rm [mas/yr]$ & 0.074 & 0.032 & 0.019 \\[2pt] 
    $ v_{z_0} \, \rm [km/s]$ & -2.455 & 0 & 5 \\[2pt] 
    \hline
\end{tabular}
\tablefoot{$\xi$ and $\sigma$ represent the mean and the standard deviation of the Gaussian distributions, respectively, which  come from \citet{2015MNRAS.453.3234P}.}

\label{priors_offset}
\end{table}

In the sampling phase of the MCMC implementation, we used 64 walkers and $10^5$ iterations. The burning-in phase was skipped and the last $80\%$ of the chains was used to compute the mean and standard deviation of the posterior distributions of the parameters. The convergence of the MCMC analysis was ensured by means of the autocorrelation time $\tau_c$, that is, we ran $N$ iterations such that $N \gg 50 \, \tau_c$.

\subsection{Results}
\label{sec:results}
We can classify three different regimes in the posterior distributions $P(|\alpha||D)$, according to the value of $\lambda$ with respect to the orbital range of S$2$, which is $1.7 \cdot 10^{13} \, \rm m \lesssim r_{s2} \lesssim 1.5 \cdot 10^{14} \, \rm m$. When $\lambda \ll r_{s2}$, the acceleration is no longer dependent on the parameter $\alpha$ and no meaningful constraints can be obtained in this regime. The small difference in the $95\%$ upper limit on $|\alpha|$ with the UCLA group resides in the different model implemented to fit the data. 

In particular,  \cite{Hees:2017aal} considered no Schwarzschild precession but an extended mass with power law distribution instead, which is absent from our work. The reason behind this choice comes from the fact that the presence of a spherically symmetric mass distribution around SgrA$^*$ has been extensively tested by the GRAVITY Collaboration \citep{GRAVITY:2018ofz, GRAVITY:2021xju, GRAVITY:2024tth}, finding a stringent upper limit of $M_{\rm ext} \lesssim 1200 \, M_{\odot}$. Taking into account this current upper limit, we decided to neglect the presence of an extended mass, as it would not alter our results.

When $\lambda \sim r_{s2}$, the best constraints on $|\alpha|$ can be obtained, determining the most stringent upper limit of $|\alpha| < 0.003$ for $\lambda = 3 \cdot 10^{13}\, \rm m
\sim 200 \, \rm AU$. This limit improves the previous estimate of \cite{Hees:2017aal}, which reported $|\alpha|<0.016$ for $\lambda = 150\, \rm AU \sim 2.2 \cdot 10^{13}\, \rm m$, stressing the importance of the precision of the GRAVITY instrument. 

Finally, when $\lambda \gg r_{s2}$, the only component left in the equations of motion is the monopolar term $M(1 + \alpha)/r$. This corresponds to a simple rescaling of the mass term and, hence, in this regime $M$ and $\alpha$ are completely degenerate and they cannot be constrained separately. To obtain the upper limit on $\alpha$ in this regime, the bounds on $M$ in Table \ref{table:priors} have been extended to $M \in (10^{-4}, 10^4) \cdot 10^6 \, M_{\odot}$, and the same bounds have been used for $\alpha$. 

A summary of the above results is reported in Figure~\ref{fig:uncertainties_alpha}, where the $95\%$ confidence interval on $|\alpha|$ as function of $\lambda$ is shown. Those confidence intervals are estimated as three times the standard deviation when the posterior distributions are normal or by computing the upper limit that corresponds to $95\%$ of the area below the curve when the distributions have different shapes. We note that GRAVITY data produce an overall improvement of roughly one order of magnitude over the entire parameter space tested.

If we assume that the gravitational interaction is mediated by a massive boson as in massive gravity theories (where $\alpha = 1$), the length scale $\lambda$ corresponds to the Compton wavelength of the particle; hence, an upper limit on the graviton's mass can be derived. 
Since $\alpha = 1$ is excluded at $95\%$ confidence level for $\lambda \lesssim 8 \cdot 10^{14}\, \rm m $, this lower bound on the wavelength, $\lambda$, can be translated into an upper limit on the graviton mass, corresponding to $m_g \lesssim 2.5 \cdot 10^{-22} \, \rm eV$.

\begin{figure}
\begin{center}
\includegraphics[width=0.45\textwidth]{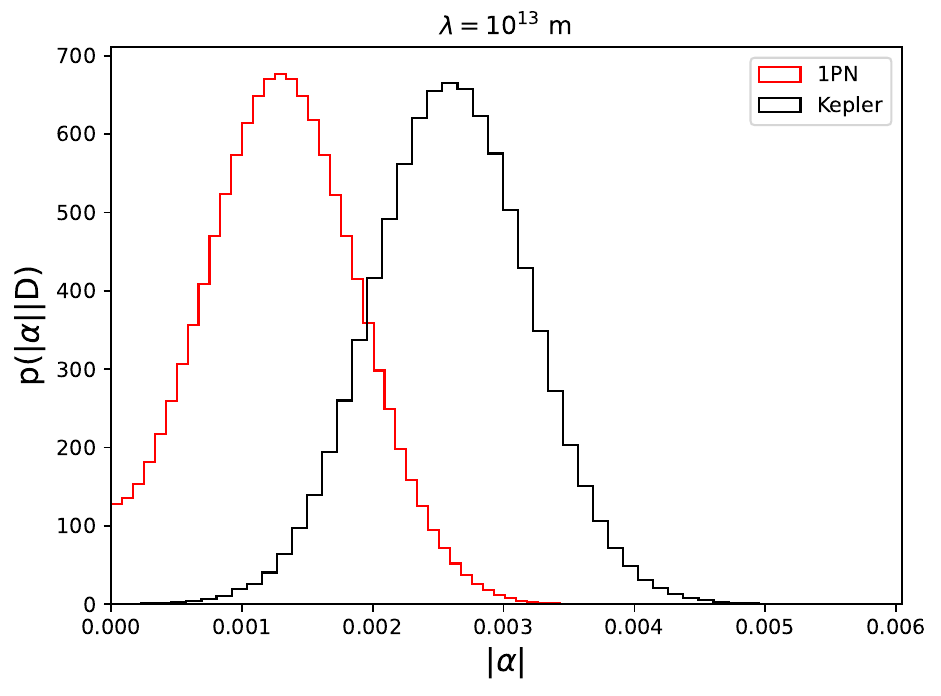} 
\end{center}
\caption{Comparison of the posterior distributions $P(|\alpha||D)$ between the Keplerian model (black curve) and  the $1$PN model (red curve). 
}
\label{fig:comparison_newton}
\end{figure}

\begin{figure}
\begin{center}
\includegraphics[width=0.45\textwidth]{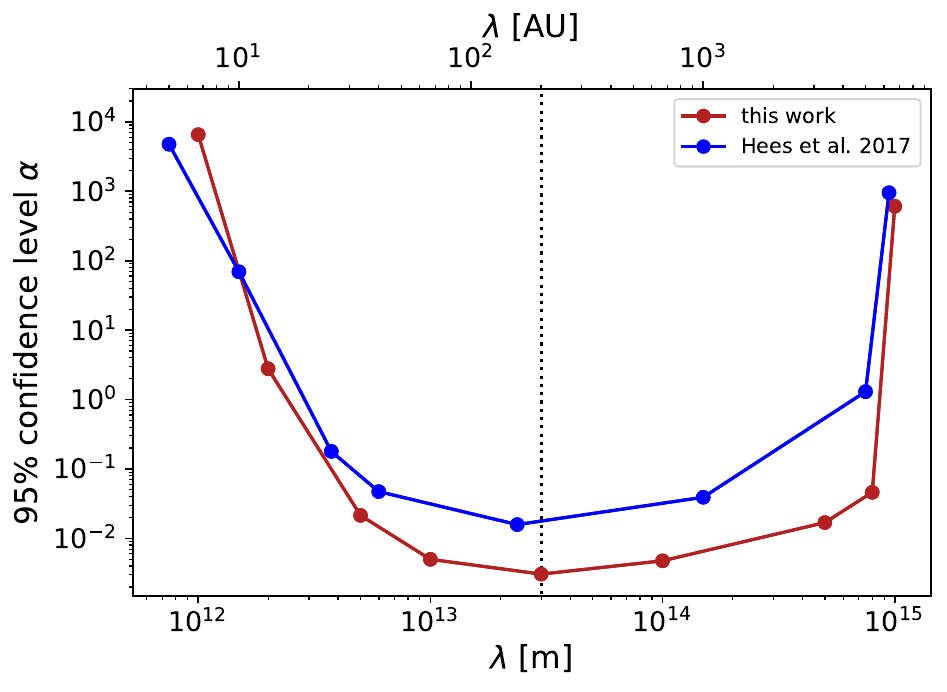} 
\end{center}
\caption{$95\%$ confidence level on $|\alpha|$ obtained in this work (red dots), compared with previous estimates by \protect\cite{Hees:2017aal} (blue dots). The dotted vertical line represents the minimum of the curve, in correspondence to $\lambda \sim 3 \cdot 10^{13} \, \text{m} \sim 200\, \text{AU}$}.
\label{fig:uncertainties_alpha}
\end{figure}

In the regime where $\lambda \sim r_{s2}$, when no $1$PN acceleration is included, the presence of the Yukawa term in the equation of motion induces a prograde precession comparable to the Schwarzschild one. This is shown in Figure~\ref{fig:comparison_newton} for $\lambda = 10^{13}\, \rm m$, where we can see that the posterior distribution $P(\alpha|D)$ is a Gaussian with mean around $\alpha \sim 0.0026$, as opposed to the $1$PN posterior.

Following \cite{Adkins:2007et}, we can compute the precession angle induced by a potential in a full orbit as 
\beq
\Delta \phi_p = - \frac{2 L}{G M e^2} \int_{-1}^{1} \frac{dz \, z}{\sqrt{1 - z^2}} \frac{dU(z)}{dz},
\eeq
where $U(z)$ is the perturbing potential evaluated at radius $r = L/(1 + e z)$ with $L= a_{\rm sma}(1-e^2)$. For $\alpha = 0.0026$, this corresponds to $\Delta \phi_p \sim 0.13^{\circ}$.
Taking the most up-to-date value of the Schwarzschild precession reported in \cite{GRAVITY:2024tth}, $\Delta \phi_{\rm Sch} = 12.1' \times (0.911 \pm 0.13) = (0.18 \pm 0.03)^{\circ}$, we can see that the precession angle induced by the Yukawa potential is compatible, within the $2 \sigma$ uncertainties, with $\Delta \phi_{\rm Sch}$.

\subsection{Comparison with theoretical estimates}
\label{subsec:comaprison_literature}

As stated in the previous section, the inclusion of the 1PN acceleration in the equations of motion implies an additional assumption; namely, that any correction to the GR expression is subleading with respect to Eq.~\eqref{1pn} and, hence, negligible in our fit. To show this, we compared our results obtained using Eq.~\eqref{1pn} with the analytical expressions derived in the literature for some specific theories. 

\cite{Alves:2023cuo} derived the 1PN acceleration in massive Brans-Dicke theory, which results in exactly the same expression as Eq.~\eqref{1pn}, as long as we identify $\alpha = a_1/2$, where $a_1 = 2 \varphi/(2 \omega_0 + 3)$, and set the reduced mass to $\eta = 0$, which is clearly a good approximation for the SgrA$^*$-S$2$ system. 

\begin{figure}
\begin{center}
\includegraphics[width=0.45\textwidth]{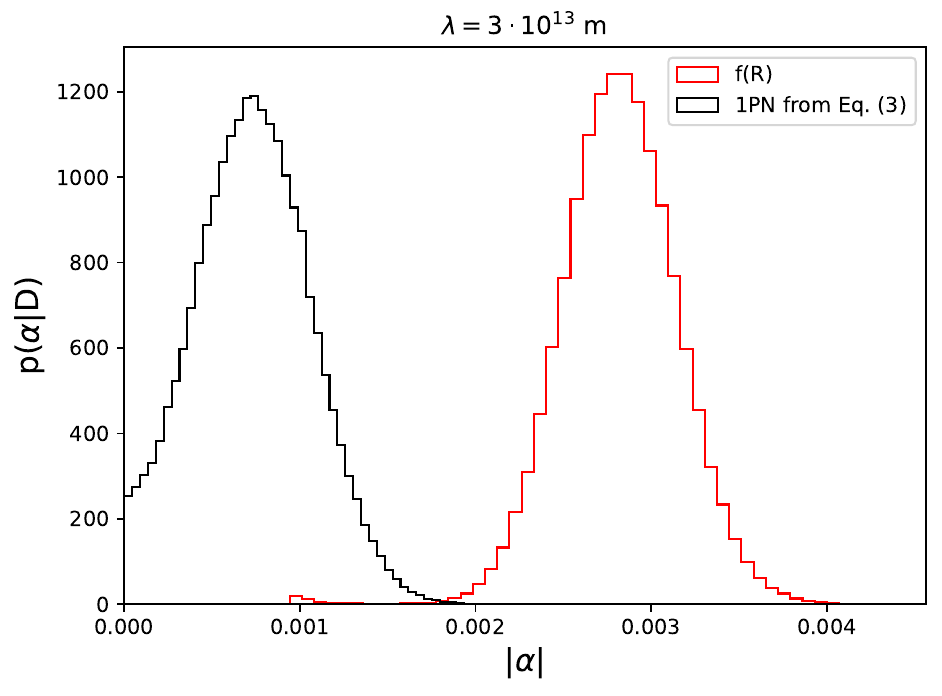} 
\end{center}
\caption{Posterior probability density of $|\alpha|$ using the $1$PN expression in Eq.~\eqref{1pn} (black curve) versus the $1$PN expansion for $f(R)$ gravity derived in \protect\cite{Tan:2024wuk} (red curve), when $\lambda = 3 \cdot 10^{13}$ m.
}
\label{fig:alpha_fR}
\end{figure}

In \cite{Tan:2024wuk} an analytical expression for the 1PN acceleration in $f(R)$ gravity is obtained and reported in their Eq.~$(17)$. We used this expression to show that our upper limits on $\alpha$ are not affected by this difference in the 1PN expansion, at least when the uncertainty on $\alpha$ is the smallest (i.e., for $\lambda = 3 \cdot 10^{13}\, \rm m)$. In Fig.~\ref{fig:alpha_fR}, the comparison between the posteriors is reported, showing that the upper limit on $\alpha$ is only changed by a factor of $2$ when the full expression for $f(R)$ gravity is used. 

\cite{Tan:2024wuk} also showed that the use of a multiple-star fit (specifically including S$2$, S$29,$ and S$55$) with the $1$PN acceleration for $f(R)$ gravity derived in Eq.~$(17)$ could potentially break the degeneracy between $M$ and $\alpha$ in the large $\lambda$ limit. This could produce a stringent upper limit on the fifth force intensity  for $\lambda > 10^{15}\, \rm m$ as well. 

In addition,  \cite{Losada:2024wdd} demonstrated that combining S$2$ motion with S$62$ lensing observations can also potentially break the degeneracy between the parametrized PN parameters $\beta$ and $\gamma$. However, we leave a multi-star analysis for a future work.

\section{Conclusions}
In this paper, we update the current constraints on the fifth force intensity at the GC using GRAVITY data for S$2$ from $2017$ to $2022$, including the pericenter passage. These data have allowed us to significantly improve previous estimates on the same effect, giving a $95\%$ confidence level curve that is one order of magnitude below the previous estimates (see Figure~\ref{fig:uncertainties_alpha}). Specifically, three different behaviors are found in the posterior distribution of $\alpha$, according to the value of the length scale of the new interaction $\lambda$ compared to S$2$ orbital range. 

The minimum value of $\alpha$ is found for $\lambda = 3 \cdot 10^{13} \, \rm m \, (\sim 200 \, AU)$ where $|\alpha|< 0.003$. 
For comparison, in correspondence with the minimum found by \cite{Hees:2017aal}, $\lambda = 150 \, \rm AU \sim 2.2 \cdot 10^{13}\, m$, we found $|\alpha|< 0.0031$. We also showed that the $1$PN expansion used in this work coincides with the expression developed for massive Brans-Dicke theory and that additional terms proportional to $\alpha$ in the $1$PN acceleration for $f(R)$ theories are subdominant with respect to the expression used in this work and thereby negligible, as the upper limit on $|\alpha|$ is unaffected. 

A complete analysis including all S-stars, specifically those with apocenter passage farther away than S$2$, is left for future work, with the aim of further improving the confidence level curve. This would allow us to possibly expand the range of $\lambda$ and obtain additional and meaningful constraints at $\lambda \gtrsim 8 \cdot 10^{14}\, \rm m$.

\begin{acknowledgements}
     A.F. and F.V. would like to thank Aurélien Hees and Laura Bernard for useful comments and fruitful discussions during the preparation of this work. We are very grateful to our funding agencies (MPG, ERC, CNRS [PNCG, PNGRAM], DFG, BMBF, Paris Observatory [CS, PhyFOG], Observatoire des Sciences de l'Univers de Grenoble, and the Funda\c c\~ao para a Ci\^encia e a Tecnologia), to ESO and the Paranal staff, and to the many scientific and technical staff members in our institutions, who helped to make NACO, SINFONI, and GRAVITY a reality. 
    This project has received funding from the European Union's Horizon 2020 research and innovation programme under the Marie Sklodowska-Curie grant agreement No 101007855.
    We acknowledge the financial support provided by FCT/Portugal through grants 
    2022.01324.PTDC, PTDC/FIS-AST/7002/2020, UIDB/00099/2020 and UIDB/04459/2020.
    J.S. acknowledge the National Science Foundation of China (12233001) and the National Key R\&D Program of China (2022YFF0503401)

\end{acknowledgements}

\bibliographystyle{aa}
\bibliography{biblio}

\appendix

\section{Details of the numerical integration}
\label{app:num_integration}

The numerical integration of the equation of motion is performed using the Python library scipy.integrate.solve\_ivp with a Runge-Kutta 5(4) algorithm, which means that the steps are evaluated using a fifth-order method, while the error is controlled assuming the accuracy of the fourth-order method. The convergence of the integration is ensured by looking at the conservation of energy over the entire integration period (almost two orbits in $\sim 30$ years gives $\Delta E/E \sim \mathcal{O}(10^{-10})$).

Kepler's equation can be solved instead using a Python root finder (scipy.optimize.newton), which implements the Newton-Raphson method. The latter solves the equation with a precision of $\mathcal{O}(10^{-16})$.

\section{Coordinates transformations and inclusion of relativistic effects.}
\label{app:relativistic_effects}

\subsection{Coordinate transformation}
\label{app:coord_transf}

The transformation from the orbital reference frame to the observer reference frame can be achieved by using the following conversion:
\beq
\begin{split}
& x' = A x_{\rm BH} + F y_{\rm BH}, \,\,\,\,\,\,\,\,\,\,\,\,\,\,\,\,\,\,\,\,\, v_{x'} = A v_{x_{\rm BH}} + F v_{y_{\rm BH}}, \\
& y' = B x_{\rm BH} + G y_{\rm BH},\,\,\,\,\,\,\,\,\,\,\,\,\,\,\,\,\,\,\,\,\, v_{y'} = B v_{x_{\rm BH}} + G v_{y_{\rm BH}}, \\
& z_{\rm obs} = -(C x_{\rm BH} + H y_{\rm BH}), \,\,\,\,\,\,\,\,\,\,\,\, v_{z_{\rm obs}} = -(C v_{x_{\rm BH}} + H v_{y_{\rm BH}}) \, ,
\end{split}
\label{coord_obs}
\eeq 
where $A, B, C, F, G, H$ are the Thiele-Innes parameters \citep{Catanzarite:2010wa} defined as:
\begin{equation}
\begin{split}
& A = \cos \Omega \cos \omega -\sin \Omega \sin \omega \cos i, \\
& B = \sin \Omega \cos \omega + \cos \Omega \sin \omega \cos i, \\
& F = -\cos \Omega \sin \omega - \sin \Omega \cos \omega \cos i, \\
& G = -\sin \Omega \sin \omega + \cos \Omega \cos \omega \cos i, \\
& C = - \sin \omega \sin i, \\
& H = -\cos \omega \sin i \,. 
\end{split}
\end{equation}
The Cartesian coordinates $\{x_{\rm BH}, y_{\rm BH}, z_{\rm BH}\}$ and velocities $\{v_{x_{\rm BH}}, v_{y_{\rm BH}}, v_{z_{\rm BH}}\}$ are those obtained from the numerical integration. For a more detailed discussion on how the coordinate system $\{x', y', z_{\rm obs}\}$ and the above transformation are defined, we refer to Figure 1 and Appendix B of \citet{Grould:2017bsw}. 

\subsection{Relativistic effects and R{\o}mer's delay}

To produce a better fit, there are observational effects that must be included in the model. 

 R{\o}mer's delay is the difference between the time of emission of the signal $t_{\rm em}$ and the actual observational dates $t_{\rm obs}$, due to the finite speed of light. To include this delay, we used the first order Taylor's expansion of the R{\o}emer equation, which is expressed as:
\begin{equation}
    t_{\rm em} = t_{\rm obs} - \frac{z_{\rm obs}(t_{\rm obs})}{1 + v_{z_{\rm obs}}(t_{\rm obs})} \,.
    \label{t_em}
\end{equation}
\noindent The difference between the exact solution of R{\o}emer equation and the approximated solution in \eqref{t_em} is at most $\sim 4$ s over the S$2$ orbit and therefore negligible. The R{\o}mer effect affects both the astrometry and the spectroscopy, with an impact of $\approx 450 \, \mu$as on positions and $\approx 50$ km/s at periastron on radial velocities. Our results recover the previous estimates for this effect reported in \cite{Grould:2017bsw, GRAVITY:2018ofz}.

Moreover, there are two relativistic effects that must be taken into account when S$2$ approaches the periastron: the relativistic Doppler shift and the gravitational redshift. Both induce a shift in the spectral lines of S$2$ that affects the radial velocity measurements. 
The former is given by
\beq 
1 + z_{D} = \frac{1 + v_{z_{\rm obs}}}{\sqrt{1- v^2}} \,,
\eeq 
while the gravitational redshift is defined as 
\beq 
1 + z_{\rm G} = \frac{1}{\sqrt{1 - 2 U(r_{\rm em})}}\,,
\eeq
where $U(r_{\rm em})$ is the potential in Eq.~\eqref{yukawa} evaluated at the time of emission $t_{\rm em}$. 

The two shifts can be combined using Eq.~(D.13) of \citet{Grould:2017bsw} to obtain the total radial velocity 
\beq
V_R \approx \frac{1}{\sqrt{1 - \epsilon}} \cdot \frac{1 + v_{z_{\rm obs}}/\sqrt{1-\epsilon}}{\sqrt{1 - v^2/(1 - \epsilon)}} - 1 \,.
\eeq
where $\epsilon = 2U({r_{\rm em}})$. 

In the total space velocity, $v = |\textbf{v}|$, we must also add a correction due to Solar System motion. We followed the most recent work of \citet{2020ApJ...892...39R} and take a proper motion of Sgr~A$^*$ of 
\beq
\begin{split}
& v_x^{\rm SSM} = -5.585 \, \rm mas/yr = 6.415 \cos(209.47^{\circ}) \, mas/yr \, ,\\
& v_y^{\rm SSM} = -3.156 \, \rm mas/yr = 6.415 \sin(209.47^{\circ}) \, mas/yr \, .
\end{split}
\eeq

\end{document}